\documentclass{INTERSPEECH2023}
\interspeechcameraready 
\usepackage{amsmath,graphicx,booktabs,multirow}
\usepackage{float}                  % 图片浮动位置
\usepackage{subfig}                 % 子图包，不要与{subfigure}混用，{subfig}较新
\usepackage{overpic}                % 叭啦叭啦，与图片排版相关

\title{Task-Agnostic Structured Pruning of Speech Representation Models}
\name{Haoyu Wang$^1$, Siyuan Wang$^1$, Wei-Qiang Zhang$^1$\sthanks{* Corresponding author}, Hongbin Suo$^2$, Yulong Wan$^2$\thanks{This work was supported by the National Natural Science Foundation of China under Grant No. 62276153.}}
\address{
  $^1$ Department of Electronic Engineering, Tsinghua University, Beijing 100084, China \\
  $^2$ Data \& AI Engineering System, OPPO, Beijing 100026, China
}
\email{w-hy21@mails.tsinghua.edu.cn, wq-zhang@tsinghua.edu.cn}

\begin{document}

\maketitle
\begin{abstract}
Self-supervised pre-trained models such as Wav2vec2, Hubert, and WavLM have been shown to significantly improve many speech tasks. However, their large memory and strong computational requirements hinder their industrial applicability. Structured pruning is a hardware-friendly model compression technique but usually results in a larger loss of accuracy. In this paper, we propose a fine-grained attention head pruning method to compensate for the performance degradation. In addition, we also introduce the straight through estimator into the $L_0$ regularization to further accelerate the pruned model. Experiments on the SUPERB benchmark show that our model can achieve comparable performance to the dense model in multiple tasks and outperforms the Wav2vec 2.0 base model on average, with 72\% fewer parameters and 2 times faster inference speed.
\end{abstract}
\noindent\textbf{Index Terms}: Model pruning, knowledge distillation, model compression, representation learning

\section{Introduction}
Recently, self-supervised pre-training has become one of the most attractive topics in the speech domain \cite{mohamed_self-supervised_2022, zhao_improving_2022}. With this method, a large amount of unlabeled data can be used to train a deep model to extract high-level representations from raw audio, which can bring significant improvement to many downstream tasks.

While pre-trained models provide a tremendous performance improvement, they also require large amount of memory and computing power. Large self-supervised pre-trained speech models such as Wav2vec2 \cite{baevski2020wav2vec}, Hubert \cite{hsu2021hubert}, and WavLM \cite{Chen2021WavLM} typically have hundreds of millions of parameters, making them unsuitable for use on consumer products such as laptops and smartphones. This is an obstacle to the application of these models in many real-world scenarios. As a result, model compression has become a major concern for these large self-supervised models.

% Some works have focused on the compression of the self-supervised pre-trained models and the commonly used methods can be divided into two categories: knowledge distillation and model pruning.

Knowledge distillation usually uses a teacher model to guide a smaller student model, and the structure of the student model must be carefully designed to achieve better performance. DistilHubert \cite{chang2022distilhubert} distills a 12-layer Hubert-based model to obtain a 2-layer student model and significantly reduces the model size. 
%They design a multitask distillation method where the hidden states of the student model are projected by a group of prediction heads and used to predict the teacher representations from higher layers.
FitHubert \cite{lee2022fithubert}, which is inspired by FitNets \cite{romero2014fitnets}, designs a thin but deep student network to provide better representation ability.
% FitHubert achieves better performance with fewer parameters than the DistilHubert model.

Model pruning attempts to discard the unimportant weights and obtain a subnetwork from the pre-trained model. In unstructured pruning, these discarded weights are randomly distributed in the matrices; in structured pruning, network units such as attention heads or feed-forward layers are removed entirely. Structurally pruned models do not require specially designed hardware for acceleration, which may be more appropriate for consumer devices. LightHubert treats model pruning as a neural architecture search problem and significantly reduces the performance degradation, but the search process still requires some time-consuming manual selections \cite{wang2022lighthubert}. Peng et al. propose a more flexible method by applying the $L_0$-regularization-based pruning method \cite{louizos2018hardtanh} to the Wav2vec 2.0 model, but their method is task-specific and comes at some additional cost when applied to downstream tasks \cite{peng_structured_2023}.

We attempt to use a similar $L_0$-regularization-based method to obtain a task-agnostic compressed model. However, learning the pruning masks using $L_0$ regularization on unsupervised pre-training tasks such as contrastive predictive coding \cite{oord2018representation} requires large computational resources. The combination of distillation and pruning is a promising solution \cite{xia2022structured,sanh2020movement}. The representation provided by the pre-trained model not only reduces the training effort of the downstream models, but also provides task-independent information for model pruning. 

Compared to existing unstructured pruning methods of the pre-trained speech models \cite{yang2022asrpathway,lai_parp_2021}, structure pruning usually suffers from a larger performance degradation \cite{liurethinking}. The crux of this problem is that using structure rather than individual weights as the basic unit of pruning reduces the degree of freedom, resulting in the removal of some important weights. To compensate for the performance degradation, we introduce a fine-grained attention head pruning method that prunes each attention head separately. To promote the pruning of coarse-grained structures and further speed up the pruned model, we also introduce the straight through estimator (STE) \cite{bengio2013ste} into the mutil-scale structured pruning method \cite{xia2022structured} based on $L_0$ regularization.

 Experiments on the SUPERB benchmark show the generalization ability of the proposed model on different downstream tasks. With the help of the pre-trained teacher, the proposed model is task-agnostic and can be directly fine-tuned to many downstream tasks. Further contrast experiments demonstrate the effectiveness of fine-grained attention head pruning and STE. Our model outperforms the distilled baselines, and achieves comparable results to the teacher model on multiple tasks, with 72\% fewer parameters and 2 times faster in speed.

% 6h-800 words
% \section{Methods}
% In this section, we will first present the foundations of our work: pre-trained speech representation models and the structured pruning method based on $L_0$ regularization. Then, we will propose a flexible fine-grained attention head pruning method, and show how the straight-through estimator brings more speedup.

\section{Backgrounds}
\subsection{Pre-trained Speech Representation Models}

Our experiment is mainly performed on WavLM \cite{Chen2021WavLM}, but the method can be easily extended to Wav2vec 2.0 \cite{baevski2020wav2vec}, data2vec \cite{baevski_data2vec_nodate}, Hubert \cite{hsu2021hubert}, and other models with similar transformer-based structures. 

WavLM is a set of state-of-the-art self-supervised pre-trained models. During pre-training, offline clustered units are used as the training target and the models learn to represent the continuous inputs by some discrete hidden units. WavLM also introduces masked speech denoising and gated relative position bias to improve the performance.

\subsection{Pruning Based on the $L_0$ Regularization}

Pruning based on $L_0$ regularization is one of the mask learning methods. In some pruning methods, parameters are discarded according to some artificially set criteria, such as the magnitude of weights or gradients. On the other hand, mask learning methods tend to consider pruning as an optimization problem \cite{louizos2018hardtanh}. As the name implies, $L_0$-regularization-based pruning adds a mask to the parameters (or parameter groups) and uses the $L_0$ norm of these pruning masks as a regularization term of the loss function. For example, in our experiments, the training objective is:
\begin{align}
    \mathcal{R}(\theta, \pi)=E_{z\sim{q(\pi)}}[\frac{1}{N}\sum^{N}_{i=1}\mathcal{L}(f_s(x_i, \widetilde{\theta}), f_t(x_i))+\lambda ||\widetilde{\theta}||_0]\text{,}
    \label{loss}
\end{align}where $f_s$ and $f_t$ are the student and teacher models for knowledge distillation, $x_i$ is the $i$th input data, $\theta$ is the parameter set of the student model, $z \in \{0,1\}$ is the pruning mask set, $\widetilde{\theta}=\theta \odot z$ is the parameter set after masking. The discrete random variable $z$ follows a Bernoulli distribution $q(\pi)$.

However, this objective function cannot be optimized by gradient descent methods because the process of sampling $z$ for $q(\pi)$ is not differentiable. Louizos et al. introduce a reparameterization trick to deal with this problem \cite{louizos2018hardtanh}. After the reparameterization, z becomes a continuous variable, determined by a learnable parameter $\alpha$ and an additional random variable $u$ that ``collects" the randomness from $z$. Formally speaking, $z$ is computed by:
\vspace{-1em}
\begin{align}
    \begin{split}
    u &\sim U(0, 1) \text{, }
    s = \text{sigmoid}(\frac{1}{\beta}\text{log}(\frac{u}{1-u}) + \text{log}\alpha) \\
    \bar{s} &= {s(\zeta-\gamma)+\gamma} \text{, }
    {z} =\text{hardtanh}({\bar{s}})\text{,}
    \label{calz}
    \end{split}
\end{align}where $u$ is sampled from a uniform distribution $U(0, 1)$, $\zeta=1.1$, $\gamma=-0.1$ are 2 constants to scale $s$ to a larger interval and make sure $z$ can be exactly 0 or 1. $\beta$ controls the temperature, and $\alpha$ is the learnable parameter. 
%After the reparameterization tick, the second term of the r.h.s of Eq. \ref{loss} has a close-from expression:

% \begin{align}
%     E_{z\sim{q(\pi)}}[||\theta \odot z||_0]=\sum^{\theta}_{j=1}\text{sigmoid}(\text{log}\alpha_{j}-\beta\text{log}-\frac{\gamma}{\zeta})
%     \label{eqsz}
% \end{align}

\begin{figure}[t]
	\label{func}
	\centering
        \subfloat[\label{pz}]{\includegraphics[width=.5\linewidth]{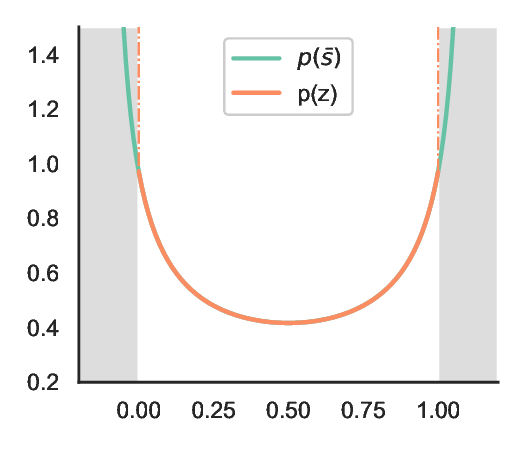}}
        \subfloat[\label{z}]{\includegraphics[width=.5\linewidth]{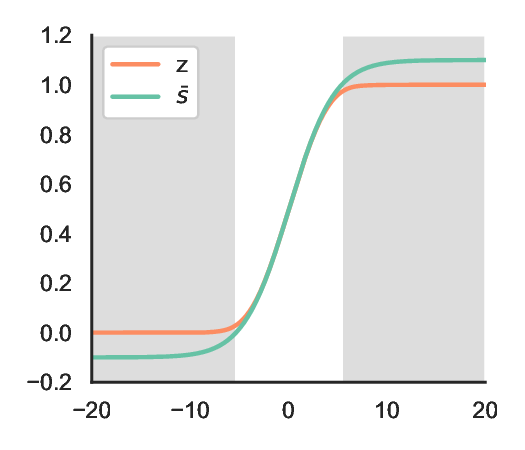}}
        \vspace{-1em}
        \caption{\textbf{(a)} the possibility distribution of $z$ and $\bar{s}$. \textbf{(b)} $z$ and $\bar{s}$ as a function of log$\alpha$, averaged on 500 samples. $z$ can be exactly 0 or 1 or any value in between. In the shadow region, $\partial z/\partial \bar{s}=0$. }
        \vspace{-2em}
 \end{figure}

Figure \ref{pz} shows the probability distribution of $z$ and $\bar{s}$, while figure \ref{z} shows their values as functions of log$\alpha$. We can see that the reparameterization trick turns the discrete masks $z$ into continuous variables while still allowing them to be exactly 0 or 1.

\subsection{Multi-scale Structured Pruning}

The $L_0$ regularization does not limit the grain of the pruning. If $z$ masks some structure, $L_0$ regularization can be used for structured pruning. The grain can be as large as an entire layer or as small as a certain dimension of a weight matrix. Recently, Xia et al. introduce a multi-scale pruning method that removes fine-grained and coarse-grained structures in parallel to promote the removal of large structures and achieve further speedup \cite{xia2022structured}. We introduced this method to increase the possibility of removing coarse-grained structures to compensate for the potential negative effects of our fine-grained attention head pruning method on the inference speed of the model.

% In their method, 4 groups of pruning masks are added to each transformer block such that:

% \begin{align}
%     \begin{split}
%         f_{\rm{MHA}}(X) &= z_{\rm{MHA}} \cdot Concat(f_{\rm{ATT}}(X))W_O \\ 
%         f_{\rm{ATT}}(X) &= z_{\rm{head}}^i \cdot (S_c(XW_V^i))^T \\
%         S_c &= \text{Softmax}((W_Q^i X)(W_K^i X)^T) \\
%         f_{\rm{FFN}}(X) &= z_{\rm{FFN}} \cdot \text{gelu}(XW_{\rm{U}}) \cdot \text{diag}(z_{\rm{int}}) \cdot W_{\rm{D}}\text{,}
%     \end{split}
%     \label{eq:att}
% \end{align}
% where $X$ is the input data, $W_Q^i$, $W_K^i$, $W_V^i$, $W_O$ is the query, key, value and output matrices, respectively. $z_{\rm{MHA}}$, $z_{\rm{head}}$, $z_{\rm{FFN}}$, $z_{\rm{int}}$ denote the pruning mask for multi-head attention layers, attention heads, feed-forward layers, and intermediate dimensions. We omit the scale factors in $f_{\rm{ATT}}(X)$ for clarity, and please note that $W_O$ should also be pruned according to $z_{\rm{head}}^i$.

\section{Methods}
\subsection{Fine-grained Attention Head Pruning}

% % Pruning masks for feed-forward layers are proved to have more influence on compression \cite{wu2020lite, ganesh2021compressing}.
In previous works \cite{peng_structured_2023,xia2022structured}, the attention heads are used as the smallest units for pruning. This may reduce the degree of freedom of pruning and lead to more performance degradation. To make structure pruning more flexible, we propose a fine-grained attention method that separately prunes each dimension of matrices in the attention layer based on the multi-scale structured pruning method of Xia et.al \cite{xia2022structured}. Formally speaking, a transformer block is masked as follows:
\vspace{-0.5em}
\begin{align}
    \begin{split}
        f_{\rm{MHA}}(X) &= z_{\rm{MHA}} \cdot \text{concat}(f_{\rm{ATT}}(X)) \\ 
        f_{\rm{ATT}}(X) &= S_c \cdot (XW_V^i) \cdot \text{diag}(z^i_{vo}) \\
        S_c &= \text{softmax}((X W_Q^i)\cdot\text{diag}(z^i_{qk})\cdot(X W_K^i)^T) \\
        f_{\rm{FFN}}(X) &= z_{\rm{FFN}} \cdot \text{gelu}(XW_{\rm{U}}) \cdot \text{diag}(z_{\rm{int}}) \cdot W_{\rm{D}}\text{,}
    \end{split}
    \label{eq:att}
\end{align} where $X$ is the input data, $W_Q^i$, $W_K^i$, $W_V^i$, $W_O$ is the query, key, value and output matrices, respectively. $z_{\rm{MHA}}$, $z^i_{qk}$ , $z^i_{vo}$, $z_{\rm{FFN}}$, $z_{\rm{int}}$ denote the pruning mask for multi-head attention layers, attention matrices, feed-forward layers, and intermediate dimensions. We omit the scale factors in $f_{\rm{ATT}}(X)$ for clarity, and please note that $W_O$ should also be pruned according to $z_{vo}^i$. For $W_Q, W_V\in\mathbb{R}^{d_{\rm{hidden}}\times d_{\rm{head}}}$, $z^i_{qk}$ and $z^i_{vo}$ will have $d_{\rm{head}}$ variables. 

% Where we replace $z_{\rm{head}}$ with 2 groups of masks, $z_{qk}$ and $z_{vo}$ while keeping $z_{\rm{MHA}}$ to preserve the ability to remove entire attention layers.
% Pruning masks are discrete and non-differential so they can not be optimized directly by gradient descent. Louizos et al. proposed an $\ell_0$ regularization method based on hard concrete distribution \cite{louizos2018hardtanh}. This Method changes the discrete masks into continuous stochastic masks. These continuous variables are very possible to be 0 or 1, but can also take values between them. They are calculated as follows

\subsection{Optimizing Pruning Masks with STE}

Although the reparameterization trick makes $z$ differentiable, the introduction of hardtanh in Eq. \ref{calz} creates a new obstacle to optimization. As shown in Figure \ref{z}, when $\text{log}\alpha$ takes a value in the shaded region, the presence of hardtanh makes $\partial z/\partial s=0$, and the learnable parameter $\alpha$ cannot be updated. That is to say, the model decides to keep a structure when $z$ is 1, but it cannot evaluate that decision.

This problem becomes more obvious for multi-scale structured pruning. Figure {\ref{z_change}} shows that the mean value of $z_{\text{FFN}}$ does not change during training, which makes multi-scale pruning ineffective. The reason may be that in the early stages of training, pruning the entire FFN layer can lead to a huge performance degradation, so $\alpha$ may be optimized to a large positive value, and difficult to update in the remaining training steps.

% We notice that large structures like attention layers are barely pruned in training and their pruning masks are usually around 1.0. In Contrast, the average of intermediate dimension masks is usually around 0.1. We think the reason might be the non-differential part of the hard concrete function. Figure \ref{ste} shows the possibility distribution of $s$ and the hardtanh function to get the continuous mask $z$. The hardtanh function, the hard concrete distribution function is not differential when $z$ is exactly 0 or 1. That is to say, the model decides to keep a structure when $z$ is 1, but it can not evaluate this decision. In the early stage of training, pruning the whole attention layer may lead to huge performance degradation, so the $\text{log}\alpha$s of layer masks might be a large positive value and hard to be updated in the remaining training steps. 

The failure to cut the coarse-scale structures will cause the sparse weight of the pruning model to be too dispersed, resulting in lower acceleration ratio. To address this problem, We apply the straight through estimator \cite{bengio2013ste} to make sure that the gradient can pass through the hardtanh function in Eq. \ref{calz}. Since the gradients from STE are not the gradients for the loss function, optimizing in this direction may 
not lead to the most accurate student and may cause instability near some local minima \cite{yin2019understanding}. 
For the stability of training, we define the gradient of STE such that:

\begin{align}
    \frac{\partial \mathcal{L}}{\partial\bar{s}} = 
    \begin{cases}
        1 \text{, if } \frac{\partial \mathcal{L}}{\partial z} >= 1\text{; } \\
        -1 \text{, if } \frac{\partial \mathcal{L}}{\partial z} < -1\text{; } \\
        \frac{\partial \mathcal{L}}{\partial z} \text{, otherwise. }
    \end{cases}
\end{align}

\subsection{Training Objective}
 Hidden states of different layers contain different types of information \cite{chang2022distilhubert,chen_optimize_2022}. Therefore, we follow Xia et al. \cite{xia2022structured} to use learnable multi-task knowledge distillation to learn the representation of different layers. We also follow Wang et al. to change the 2nd term on the r.h.s of eq. \ref{loss} into  a Lagrangian term to better control the sparsity \cite{wang2020largs}. Our training objective is as follows: 
\begin{align}
        \mathcal{L}=\frac{1}{N}\sum_{i=0}^{N}\sum_{(j,k)\in D}\mathcal{L}_{\rm{MSE}}(h_i^j, \hat{h}_i^k)+\lambda_1(\hat{p}-p) + \lambda_2(\hat{p}-p)^2\text{,}
\end{align}where $\hat{p}$ is the approximate model sparsity, $p$ is the target sparsity. $\lambda_1$ and $\lambda_2$ are learnable parameters for the Lagrangian regularization. $D$ is the teacher-student layer pairing relation learned during training \cite{xia2022structured}, for sample $i$, $h_i^j$ and $\hat{h}_i^k$ are the output of layer $j/k$ of the
student and teacher models, respectively.
%  According to previous research, Distillation can help improving the performance of pruning\cite{xia2022structured,sanh2020movement,lagunas2021block}. Inspired by the CoFi pruning, we also use the dynamic layer-wise distillation. Figure \ref{loss} is an overview of our method. During training, some student layers might be removed, so we choose the 3rd, 6th, 9th, and 12th layers of the WavLM model as teacher layers, and each will dynamically select the most similar student layer to calculate the distillation loss. Different of the Cofi pruning, we find that the last hidden layer of the teacher model always selects theire student from the middle (e.g., the 10th layer). Considering that the last layer are usually selected as the feature extractor, we decided to prune all the layers after the final student.  The final loss is as follows: 
% \begin{align*}
%     L = L_{\rm{distil}} + {,}
% \end{align*}where $L_{\rm{distil}}$ is the layer-wise distillation loss, $\hat{s}$ is the approximate model sparsity, and $t$ is the target sparsity. $\lambda_1$ and $\lambda_2$ are learnable parameters.

% \begin{figure}[htb]
% \begin{minipage}[b]{1.0\linewidth}
%   \centering
%   \centerline{\includegraphics[width=7.0cm]{loss.jpg}}
% %   \centerline{(a) Result 1}\medskip
% \end{minipage}
% \caption{An overview of our layer-wise distillation. Layers after the final student are removed.}
% \label{loss}
% \end{figure}

% 6h-800 words
\vspace{-1em}
\section{Experiments}
\subsection{SUPERB}
SUPERB (Speech processing Universal PERformance Benchmark) is a benchmark for evaluating the performance of speech pre-training models \cite{yang2021superb}. SUPERB provides 10 predefined speech tasks from different perspectives where the pre-trained models are used as upstream feature extractors. These tasks include phoneme recognition (PR), automatic speech recognition (ASR), keyword spotting (KS), query-by-example spoken term detection (QbE), speaker identification (SID), automatic speaker verification (SV), speaker diarization (SD), intent classification (IC), slot filling (SF), and emotion recognition (ER).

\subsection{Pruning setup}
\textbf{Model}. Our model is initialized from the WavLM base model, which consists of a 7-layer CNN feature extractor and a 12-layer transformer encoder. For the matrices in Eq. \ref{eq:att}, $W_Q^i, W_K^i, W_V^i \in \mathbb{R}^{768\times64}$, $W_O \in \mathbb{R}^{768\times768}$, $W_U \in \mathbb{R}^{768\times3072}$, and $W_D \in \mathbb{R}^{3072\times768}$. For each transformer block, we have 12 attention heads, leading to $12*64=768$ elements in $z_{qk}$ and $z_{vo}$. We also have 3072 elements in $z_{\rm{int}}$ for each dimension in the FFN layer, and 1 element in $z_{\rm{MHA}}$ and $z_{\rm{FNN}}$ to mask the entire layer. The target pruning sparsity is set to 80\%. The teacher model of knowledge distillation is also the WavLM base model. 

\textbf{Data}. We use the 960 hours Librispeech \cite{panayotov2015librispeech} corpus for pruning. For SUPERB tasks, we use the dataset according to the official guidelines\footnote{https://github.com/s3prl/s3prl/}. 

\textbf{Pruning}. Pruning is performed on an RTX 3090 GPU for 200k steps and takes about 36 hours. Our training hyperparameters are chosen according to DistilHuBERT \cite{chang2022distilhubert} and Xia et al. \cite{xia2022structured}. The learning rate increases linearly to 2.0e-4 in the first 7\% steps and decreases linearly to 0 in the remaining steps, and the target sparsity increases linearly to 80\% in the first 7\% steps and remains constant for the rest.

\begin{table*}[!ht]
    \centering
      \caption{Results on SUPERB of the proposed model, and other baselines. The performances are evaluated by Phoneme Error Rate (PER\%), Accuracy (Acc\%), Word Error Rate (WER\%), Maximum Term Weighted Value(MTWV), F1 Score (F1\%), Concept Error Rate (CER\%), Equal Error Rate (EER\%), and Diarization Error Rate (DER\%). DistilWavLM is our reproduction of DistilHubert with the teacher changed to WavLM base; FAHP is the abbreviation for the proposed Fine-grained Attention Head Pruning method.}
    \begin{tabular}{lccccccccccc}
    \toprule
        \multirow{2}{*}{\textbf{Method}} & \textbf{KS} & \textbf{IC} & \textbf{PR} & \textbf{ASR} & \textbf{ER} & \textbf{QbE} & \textbf{SF} & \textbf{SID} & \textbf{SV} & \textbf{SD} & \multirow{2}{*}{\textbf{$\text{Superb}_s$↑}} \\ \cline{2-11}
         & Acc↑ & Acc↑ & PER↓ & WER↓ & Acc↑ & MTWV↑ & F1↑/CER↓ & Acc↑ & EER↓ & DER↓ & \\ 
        \midrule
        \multicolumn{12}{l}{\textbf{Baselines}} \\
        \midrule
        % FBANK & 41.38  & 9.65  & 82.01  & 23.18  & 48.24  & 0.58  & 69.64/52.94 & 20.06  & 9.56  & 10.05 & 0 \\ 
        wav2vec \cite{schneider_wav2vec_2019}  & 95.59  & 84.92  & 31.58  & 15.86  & 59.79  & 4.85  & 76.37/43.71 & 56.56  & 7.99  & 9.90 & 491.59 \\ 
        w2v2 Base  & 96.23  & 92.35  & 5.74  & 6.43  & 63.43  & 2.33  & 88.3/24.77 & 75.18  & 6.02  & 6.08 & 735.00 \\ 
        HuBERT Base & 96.30  & 98.34  & 5.41  & 6.42  & 64.92  & 7.36  & 88.53/25.2 & 81.42  & 5.11  & 5.88 & 837.63 \\ 
        WavLM Base  & 96.79  & 98.63  & 4.84  & 6.21  & 65.94  & 8.70  & 89.38/22.86 & 84.51  & 4.69  & 4.55 &  895.99 \\ 
        \midrule
        \multicolumn{12}{l}{\textbf{Distilled Models}} \\ 
        \midrule
        DistilHuBERT  & 95.98  & 94.99  & 16.27  & 13.37  & 63.02  & 5.11  & 82.57/35.39 & 73.54  & 8.55  & 6.19 & 647.88 \\ 
        DistilWavLM  & 96.40  & 96.39  & 14.18  & 13.24  & 63.69  & 7.07  & 85.27/31.80 & 71.00  & 8.87 & 7.2 & 668.39 \\ 
        \midrule
        \multicolumn{12}{l}{\textbf{Ours}} \\
        \midrule
        Proposed  & 96.57  & 98.08  & 9.09  & 10.61  & 63.61  & 7.40  & 87.14/27.13 & 74.56  & 6.17  & 6.11 & \textbf{769.62} \\ 
        w/o FAHP & 96.14  & 98.05  & 10.51  & 11.83  & 63.78  & 5.19  & 85.57/30.91 & 70.03  & 6.12  & 7.18 & 721.79 \\ 
        \bottomrule
    \end{tabular}
    \vspace{-1em}
    \label{main}
\end{table*}

\section{Results}
% \subsection{Evaluation on SUPERB}
Table \ref{main} shows the evaluation results on the SUPERB downstream tasks. Our model has comparable performance to the teacher model in KS, IC, ER, SV, and SD tasks, demonstrating the effectiveness of our approach. The performance degradation occurred mainly in PR, ASR, and SF tasks. These tasks require more complex content-related information, which is more likely to be lost during pruning. Using the same WavLM base teacher model, our method outperforms the distilled models in most tasks, especially in content-related tasks such as ASR, showing that our model better preserves the performance of the teacher model.

In addition to the task-specific metrics, we also use the SUPERB score ($\text{superb}_s$) to provide an overall evaluation. The SUPERB score is an average of the linear transformations of all the task-specific metrics, and is determined by the SOTA model on the benchmark and a predefined FBANK baseline. At the time of writing, the SOTA model is WavLM-Large\footnote{The performance of the WavLM-Large model can be found at https://superbbenchmark.org/leaderboard.}. Formally speaking, the SUPERB score is defined as: 

\vspace{-0.5em}
\begin{align}
    \text{superb}_s = \frac{1}{T}\sum_{t\in T}\frac{1000}{m_t^{\text{sota}}-m_t^{\text{fbank}}}(m_t^u-m_t^{\text{fbank}})\text{,}
\end{align}where $m_t^u$ is the metric of task $t$ and model $u$, $\text{superb}_s(\text{sota})\equiv1000$, $\text{superb}_s(\text{fbank})\equiv0$.

\begin{figure}[t]
    \centering
    \includegraphics[width=0.75\linewidth]{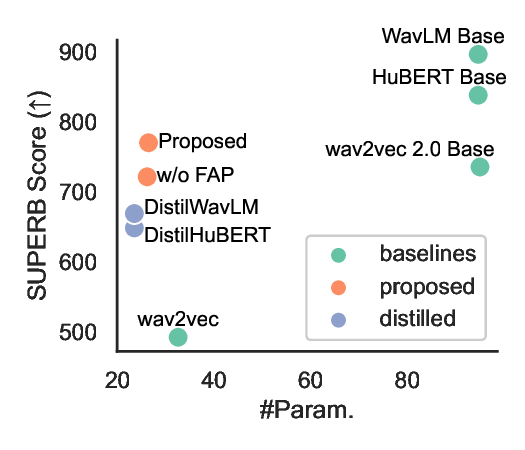}
    \vspace{-1em}
    \caption{The relationship between the SUPERB score and the
number of parameters. }
    \vspace{-1em}
    \label{fig:superbs}
    \vspace{-1em}
\end{figure}

Figure \ref{fig:superbs} shows the relationship between the SUPERB score and the
number of parameters. Our model significantly outperforms the distillation models with similar number of parameters, and even has superior performance to the Wav2vec 2.0 base model. These results show that the proposed method achieves a better balance between performance and the number of parameters compared to the distillation-based method. 

We also compare our method with the previous pruning method which directly removes the attention heads (w/o FAHP in Table \ref{main}). Again, the improvement is mainly reflected in tasks such as ASR, suggesting that fine-grained attention head pruning can help compensate for the loss of complex information in structured pruning. 

\begin{figure}[t]
    \label{ste}
    \centering
    \subfloat[The average value of $z_{\rm{FFN}}$ and $z_{\rm{MHA}}$.\label{z_change}]{\includegraphics[width=0.8\linewidth] {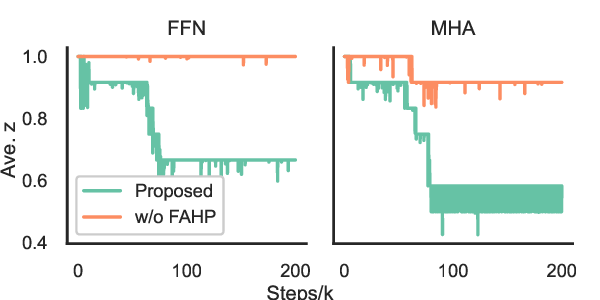}} \\
    \vspace{-0.5em}
    \subfloat[Remaining (blue) parameters in $W_V^0$ for 12 layers.\label{mat}]{\includegraphics[width=.8\linewidth]{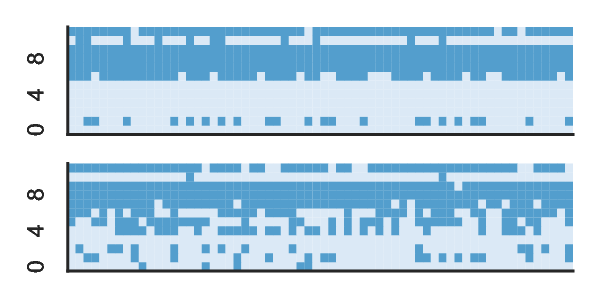}}
    \caption{The effectiveness of STE}
    \vspace{-1.5em}
 \end{figure}

Figure \ref{z_change} shows the average of the pruning masks $z_{\text{FFN}}$ and  $z_{\text{MHA}}$ during pruning. By introducing STE, the pruning masks of coarse-grained structures change more frequently and eventually drop to lower values, which proves the effectiveness of STE. Figure \ref{mat} shows the distribution of the remaining weights of each layer after pruning. Since the coarse-grained structures can be entirely removed, the remaining parameters tend to be concentrated, leading to further acceleration. 

\begin{table}[t]
    \centering
    \caption{Inference time measured on a RTX3090 GPU, by extracting features of librispeech dev-clean set and are averaged on 5 runs.}
    % \hspace*{\fill} \\
    \vspace{-0.5em}
    \begin{tabular}{lcc}
        \toprule
        \multirow{2}*{\textbf{Method}} & \textbf{\#Params} & \textbf{Infer. time} \\
        \cline{2-3}
         & Millions & Seconds \\
        \midrule
        WavLM base & 94.70  & 91.87(1.0x) \\
        Proposed & 26.57  & \textbf{46.08(1.99x)} \\
        w/o STE & 26.37 & 67.78(1.35x) \\
        \bottomrule
    \end{tabular}
    \vspace{-1em}
    \label{speed}
\end{table}

\begin{table}[t]
    \centering
    \caption{Influence of STE on accuracy. ASR, IC, ER, SID are representative of SUPERB content, paralinguistic, speaker, and semantic tasks.}
    % \hspace*{\fill} \\
    \vspace{-0.5em}
    \begin{tabular}{lcccc}
        \toprule
        \multirow{2}*{Methods} & \textbf{ASR} & \textbf{IC} & \textbf{ER} & \textbf{SID} \\
        \cline{2-5}
         & WER$\downarrow$ & Acc$\uparrow$ & Acc$\uparrow$ & Acc$\uparrow$\\
        \midrule
        proposed & \textbf{10.29} & \textbf{98.08} & 63.61 & 74.56 \\ 
        w/o STE & 10.61 & 97.07 & \textbf{64.17} & \textbf{74.65} \\
        \bottomrule
    \end{tabular}
    \vspace{-1.5em}
    \label{performance}
\end{table}

In addition, the remaining weight is concentrated at the top of the network. Since content-related information is more prominent in the features of the top layers, this distribution of remaining weights may be one of the reasons for the network's improvement in content-related tasks.

We also measure the inference time of the 2 models above. Table \ref{speed} shows the speed effect of STE. It can be seen that the concentrated weight distribution brought by STE significantly improves the inference speed of the model. With STE, the pruned model is 1.4 times faster with a similar number of parameters.

Furthermore, we show the effect of STE on accuracy. Among these 4 tasks, STE brings improvement in ASR and IC, while causing degradation in ER and SID, but both the positive and negative influence are not significant. The degradation in ER and SID may be due to the parameters removed from the lower layers that are related to speaker or emotion information.

\vspace{-0.5em}
\section{Conclusion}
% \vspace{-0.5em}
In this paper, we present a task-agnostic structured pruning method of pre-trained speech representation models. By using fine-grained attention head pruning, we retain the ability to represent content-level information and reduce the performance degradation caused by structured pruning. We introduce STE to multi-scale structured pruning to further accelerate the model. Our experiments prove that the proposed model reduces 72\% of the parameters while having comparable performance to the dense model in multiple tasks, and outperforms the Wav2vec2 base model in average performance.

\clearpage
% \section{Acknowledgements}

% This work was supported by the National Natural Science Foundation of China under Grant No. U1836219 and  No. 62276153.

% Below is an example of how to insert images. Delete the ``\vspace'' line,
% uncomment the preceding line ``\centerline...'' and replace ``imageX.ps''
% % with a suitable PostScript file name.
% % -------------------------------------------------------------------------
% \begin{figure}[htb]
% \begin{minipage}[b]{1.0\linewidth}
%   \centering
%   \centerline{\includegraphics[width=8.5cm]{image1}}
%   \centerline{(a) Result 1}\medskip
% \end{minipage}
% \caption{Example of placing a figure with experimental results.}
% \label{fig:res}
% \end{figure}

\bibliographystyle{IEEEtran}
\bibliography{refs}

\end{document}